\documentclass{caosp306}

\usepackage{graphicx}

\usepackage{natbib}
\def\BibTeX{{\rm B\kern-.05em{\sc i\kern-.025em b}\kern-.08em
             T\kern-.1667em\lower.7ex\hbox{E}\kern-.125emX}}

\begin{document}

%
\hauthor{D. Ko{\c c}ak, T. {\.I}{\c c}li and K. Yakut}

\title{Photometric study of close binary stars in the M35, M67, and M71 Galactic clusters}


%
\author{
       Dolunay Ko{\c c}ak \inst{1} 
      \and 
        Tu\u{g}\c{c}e {\.I}{\c c}li \inst{1}   
      \and 
        Kadri Yakut \inst{1}
       }

\institute{
           Department of Astronomy and Space Sciences, University of Ege, 35100, Bornova--{\.I}zmir, Turkey, \email{dolunay.kocak@gmail.com}
          }

\date{October 30, 2019}

\maketitle

\begin{abstract}
We obtained new multicolour photometry of close binary stars in the young open cluster 
M35, the solar-age open cluster M67, and the globular cluster M71. New observations have been carried out at the T\"{U}B\.{I}TAK National Observatory (TUG) by using the 100\,cm (T100) telescope. We present observational results for eclipsing 
binary systems in the selected Galactic clusters. New accurate light
curves for 2MASS  J19532554 + 1851175, 
2MASS  J19533427 + 1844047, 2MASS  J06092044 + 2415155, and AH Cnc were obtained. 
The light curves were analysed and we derived some of the orbital parameters of the systems.

\end{abstract}

\section{Introduction}
\label{intr}
Stellar clusters are very important tools for studying stellar formation and evolution, 
as well as the formation, structure, and dynamical evolution of the
Galaxy. There are more than three thousand open clusters and about a
hundred and fifty globular clusters in the Milky Way. Globular
clusters are important in determining the lower limit  of the age of
the Universe. Open clusters are made up of relatively young stars,
while globular clusters are composed of very old and metal-poor
stars. Typical globular and open clusters contain stars with very
different masses and with different properties in the HR diagram. The
binary systems in the star clusters, especially at the turn off point
of the main sequence, provide great opportunitboldies for studying the
evolution of both the clusters and the binary systems. Therefore star
clusters are ideal laboratories for testing and calibrating stellar
evolution theories. For details, see Meynet et al. (1993),
Harris (1996), Elmegreen \& Efremov (1997), Chantereau et al.(2015), Chantereau et al.(2016), 
Hurley et al. (2005),Prantzos \& Charbonnel (2006), Decressin et al. (2007), 
Yakut et al. (2009), Bilir et al. (2012), Yakut et al. (2015).

Galactic globular clusters are compact and old systems which contain
more than 1 million stars. So, there is a very high probability of
collision of the stars with each other. Globular clusters host a lot
of binary stars. Studying binaries in a cluster has some
advantages. For instance, all binaries are at equal distances from us,
have (almost) same chemical composition and (almost) same
age. Nevertheless, the masses of the stars differ from one to
another. By using stellar parameters of  binary system, we can test
the theoretical evolutionary models and elucidate some poorly
understood astrophysical phenomena such as mass loss, mass transfer, 
physical parameter variations during the evolution, angular momentum problem, etc.

The distance to the M71 globular cluster is about 4 kpc and it is
fairly metal-rich, low-density globular cluster in the Galaxy
(Grundahl et al. 2002).  M71 is an important laboratory for studying
the formation of exotic objects such as blue strugglers, cataclysmic
variables, low-mass X-ray binaries and millisecond pulsars (Ferraro et
al. 1997; Pooley et al. 2003; Heinke et al. 2005). NGC 6791 is the
oldest open cluster in the Galaxy with an age of 7.7$\times$10$^9$ years (Yakut et al., 2015). 
Be 17 and NGC 188 are also among the oldest galactic clusters (Phelps 1997; Meibom et al., 2009). NGC 2168 (M35), 
classified by Trumpler in 1930, is a rich open cluster of almost 180 Myr  old and its distance is about 900 pc. 
The open cluster M67 (NGC 2682) is located at a distance of 840 pc. This cluster has many different types of binary stars.
The cluster is also important because of its solar age and solar-like chemical composition (Yakut et al., 2009).
The chemical abundance and age of the cluster is very close to those
of the Sun, which is important for testing stellar evolution
models. M67 includes many blue stragglers that are bluer and brighter
than the stars at the turn-off point of the cluster. Moreover, the
cluster contains close and interacting binary stars like AH Cnc, ES
Cnc, and EV Cnc. 


\section{New observations}
We have obtained high precision new multi-colour observations of close binary systems in the open cluster M35, M67, and 
in the globular cluster M71. New observations were obtained using the
100 cm telescope at the T\"{U}B\.{I}TAK National Observatory (TUG). 
M71 was observed during 15 nights in the $V$ and $R$ filters. M35 is
scattered over an area of the sky almost the size of the full
moon. Therefore, we observed it by dividing the CCD field into four
regions. New CCD observations of M35 and M67 in the $V$, and $R$ filters were obtained on 12 and 4 nights, respectively.

\begin{figure}
\centerline{\includegraphics[width=7.8cm,clip=]{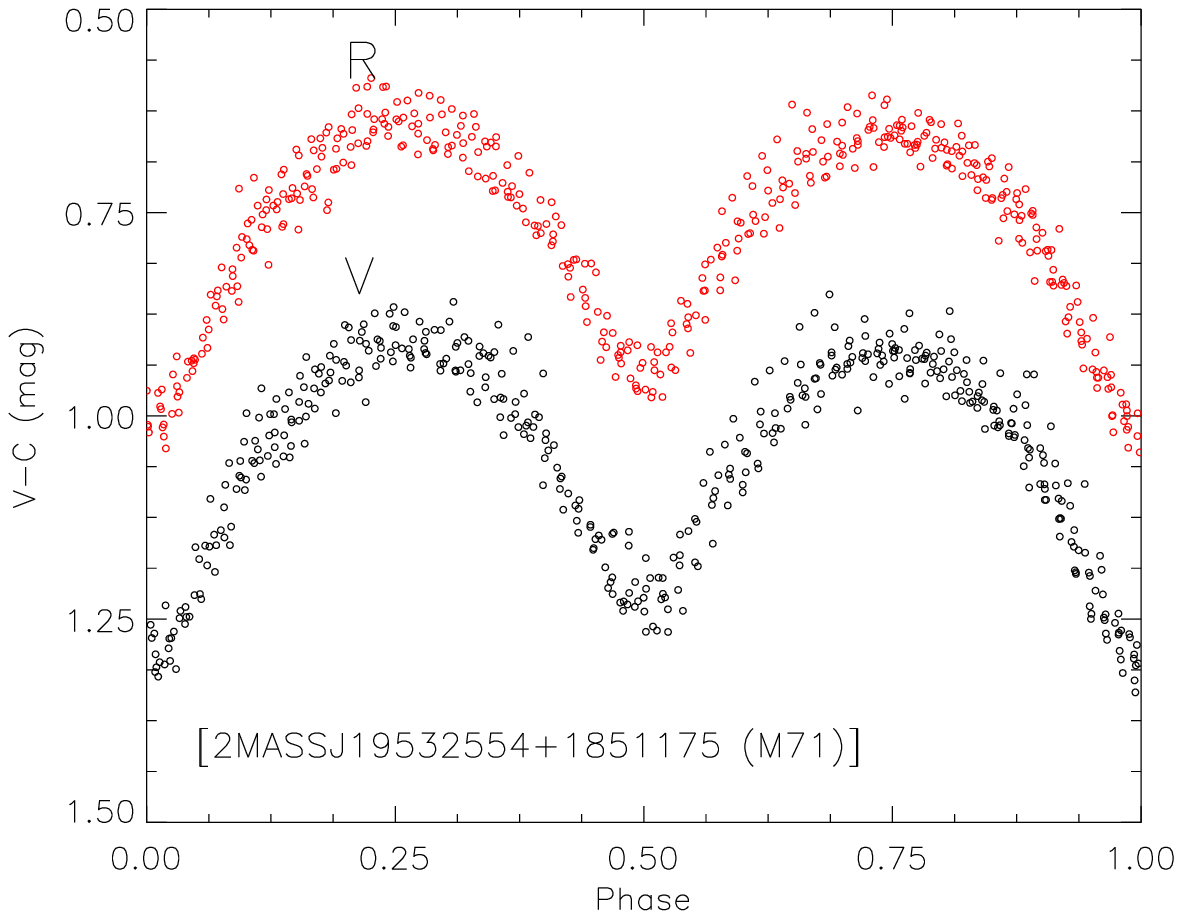} 
   \includegraphics[width=7.8cm,clip=]{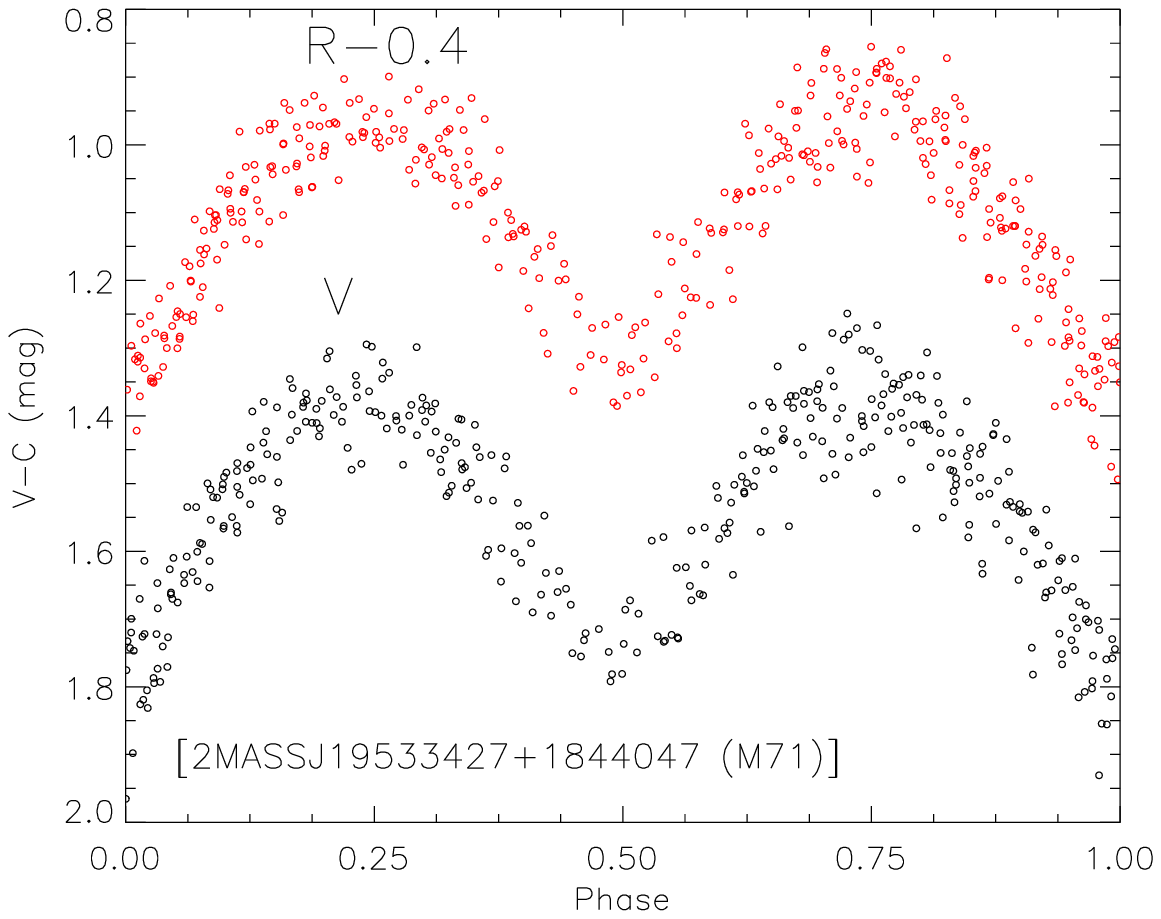}}
\centerline{\includegraphics[width=7.8cm,clip=]{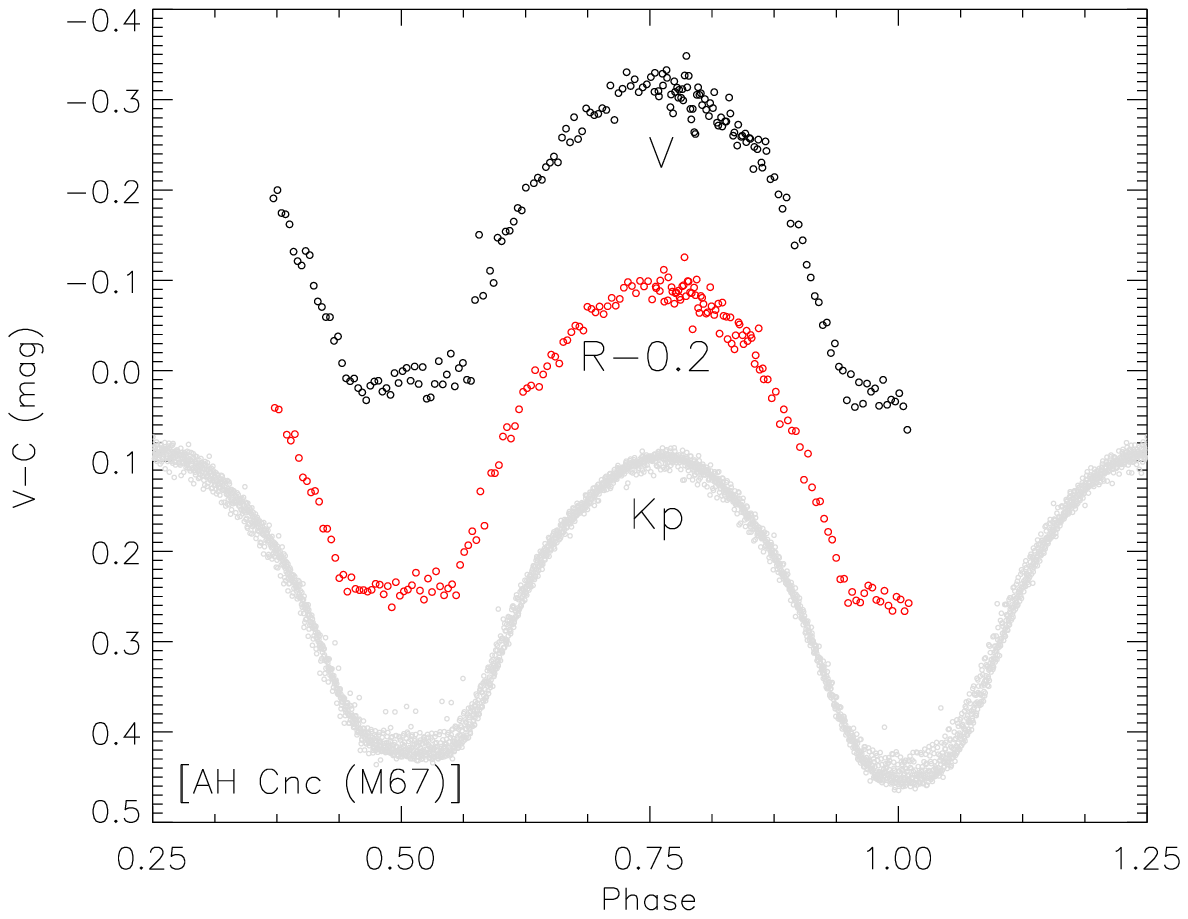} 
  \includegraphics[width=7.8cm,clip=]{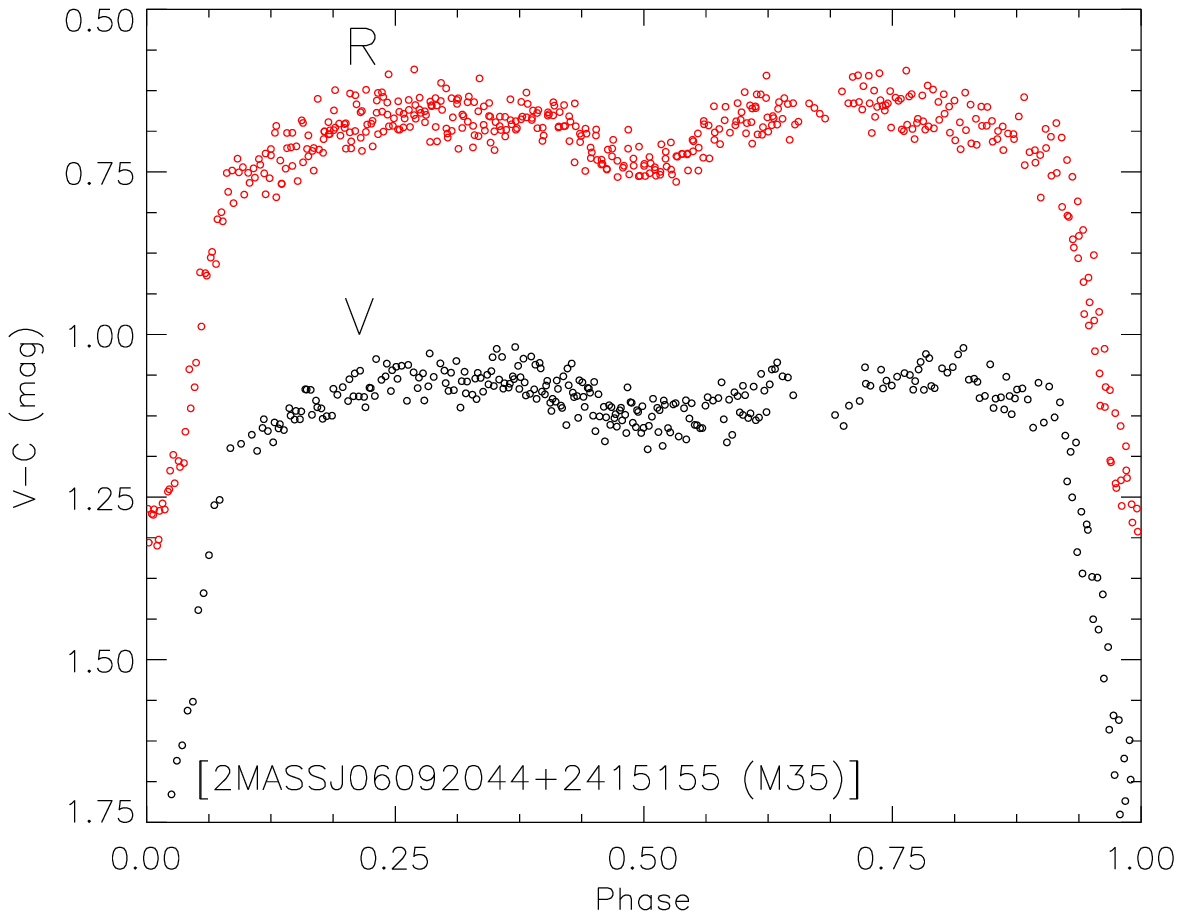}}
	\caption{$V$, $R$ and Kepler ($K_p$) light curves of some close binaries in the Galactic clusters  M71, M67, and M35.}
	\label{figure}
\end{figure}

Data reductions were performed by subtracting the bias and dark frames and dividing by the flat. 
We studied each night separately following the time correction, differential photometry was performed as we did in our earlier study ({\.I}{\c{c}}li et al., 2013). In the data reduction we used IRAF/DAOPHOT and AstroImageJ (Collins et al., 2017). 
We used {\it Kepler} satellite observations for the contact binary system AH Cnc. This raw data show some 
fluctuations due to common instrumental effects \citep{2010ApJ...713L..87J}. We eliminate these systematic variations applying  cotrending and detrending process as we did in our earlier {\it{Kepler}} studies ({\c{C}}okluk, et al., 2019; Yakut et al., 2015)
Fig.~\ref{figure} shows the light variation of the selected systems in the
galactic clusters M35 (2MASS J06092044+2415155),  M67 (AH~Cnc), and  M71 (2MASS J19533427+1844047, 2MASS J19532554+1851175).

\section{Results}

In this study, accurate multicolor light variations of some eclipsing binary systems in several galactic open and 
globular clusters were obtained. Using new observations, synthetic light curves were modelled with Phoebe ( Pr{\v{s}}a, \& Zwitter, 2005; Wilson \& Devinney, 1971;  Wilson, 1979).
During the light curves analysis, the  limb  darkening  coefficients (from van Hamme, 1993), albedos (from Rucinski, 1969) and the values of the gravity-darkening coefficients (from Lucy 1967) were taken as fixed parameters.  
Preliminary analysis resulted in determination of the orbital
inclination ($i$), the mass ratio ($q$), and the fractional radii of the primary ($r_1$) and secondary ($r_2$) components for 
the selected binaries listed in Table~\ref{table}.

\begin{table*}[t]
	\begin{center}
		\caption{Light curve solution and their formal 1$\sigma$ errors for 2MASS J06092044 + 2415155, AH~Cnc, 2MASS J19533427 + 1844047, and 2MASS J19532554 + 1851175.}
		\label{table}
		\begin{tabular}{lllll}
			\hline\hline
			System    			 & J06092044(M35)		& AH Cnc(M67)			    &	J19533427(M71)		& J19532554(M71)	\\
			\hline
			$i$    (deg) 		 & 69.1$\pm$0.7 		& 87.9$\pm$0.2 	  			& 71.8$\pm$0.2 			& 71.2$\pm$0.4 \\
			$q$    (M$_2$/M$_1$) & 0.71$\pm$0.01  		& 0.147$\pm$0.004   		& 0.13$\pm$0.01 		& 0.22$\pm$0.01  \\
			r$_1$  (R$_1$/a)     & 0.3022 $\pm$ 0.0013  & 0.5660 $\pm$ 0.0005 		& 0.567 $\pm$ 0.015     & 0.333 $\pm$ 0.008  \\
			r$_2$  (R$_2$/a)     & 0.3482 $\pm$ 0.0013  & 0.2483 $\pm$ 0.0008 		& 0.232 $\pm$ 0.047 	& 0.424 $\pm$ 0.007   \\
			\hline\hline
		\end{tabular}
	\end{center}
\end{table*}

\acknowledgements
We are grateful to Walter van Hamme for his comments and suggestions.
This study was supported by Turkish Scientific and Research Council (T\"{U}B\.{I}TAK 117F188) and the T\"{U}B\.{I}TAK National Observatory (18CT100-1422). DK thanks T\"{U}B\.{I}TAK for his Fellowship (2211-C and 2214). KY would like to acknowledge the contribution of  COST (European Cooperation in Science and Technology)  Action CA15117 and CA16104.

\end{document}